\documentstyle[sprocl,epsfig]{article}
\def\Journal#1#2#3#4{{#1} {\bf #2}, #3 (#4)}

\def\PLB{{\em Phys. Lett.}  B}

\def\PRD{{\em Phys. Rev.} D}

\def\EPJ{{\em E. Phys. J.} C}

\def\be{\begin{equation}}
\def\ee{\end{equation}}
\def\bea{\begin{eqnarray}}
\def\eea{\end{eqnarray}}
\def\Z0{$\mathrm{Z}$}
\def\MZ{$\mathrm{M}_{\mathrm{Z}}$}
\def\MW{$\mathrm{M}_{\mathrm{W}}$}
\def\MH{$\mathrm{M}_{\mathrm{H}}$}
\def\VJK{$\mathrm{V}_{\mathrm{jk}}$}
\def\mt{$\mathrm{m}_{\mathrm{t}}$}
\def\GM{$G_{\mu}$}
\def\als{$\alpha_{\mathrm{s}}$}
\def\SEF{$\sin^2\theta_{\mathrm{eff}}$}
\def\GZ{$\Gamma_{\mathrm{Z}}$}
\def\Gl{$\Gamma_{\mathrm{l}}$}
\def\Gh{$\Gamma_{\mathrm{h}}$}
\def\Rb{$\mathrm{R}_{\mathrm{b}}$}
\def\Rc{$\mathrm{R}_{\mathrm{c}}$}
\def\Rl{$\mathrm{R}_{\mathrm{l}}$}
\def\S0{$\sigma^0_{\mathrm{f\bar{f}}}$}
\def\ALR{$\mathrm{A}_{\mathrm{LR}}$}
\begin{document}

\title{PRECISION TESTS OF THE ELECTROWEAK INTERACTIONS AT $e^+e^-$ COLLIDERS}

\author{F.TEUBERT}

\address{European Laboratory for Particle Physics (CERN), 
\\ CH-1211 Geneva 23, Switzerland\\E-mail: frederic.teubert@cern.ch}


\maketitle\abstracts{ This paper is an updated version of the 
lectures given at the XXIX International Meeting on Fundamental Physics 
in Sitges, Barcelona (February 2001).
The measurements performed at LEP and SLC 
have substantially improved the precision of the test of the 
Minimal Standard Model. The precision is such that there 
is sensitivity to pure weak radiative corrections. This allows
to indirectly determine the top mass (\mt=180$\pm$10~GeV),  
the W-boson mass (\MW=80.375$\pm$0.022~GeV), and to set an upper limit on the
the Higgs boson mass of 196~GeV at 95\% confidence level. }

\section{\bf Introduction}

In the context of the Minimal Standard Model (MSM), any ElectroWeak (EW) process 
can be computed at tree level from $\alpha$ (the fine structure constant 
measured at values of $q^2$ close to zero), \MW~(the W-boson mass), 
\MZ~(the Z-boson mass), and \VJK~(the 
Cabbibo-Kobayashi-Maskawa flavour-mixing matrix elements).  

When higher order corrections are included, any observable can be predicted in the 
``on-shell'' renormalization scheme as a function of:

\noindent
\bea
O_i & = & f_i(\alpha, \alpha_{\mathrm{s}},\mathrm{M}_{\mathrm{W}} ,\mathrm{M}_{\mathrm{Z}},
\mathrm{M}_{\mathrm{H}},\mathrm{m}_{\mathrm{f}},\rm{V}_{\rm{jk}})  \nonumber 
\eea

\noindent and contrary to what happens with ``exact gauge symmetry theories'', 
like QED or QCD, the effects of heavy particles do not decouple. Therefore, the 
MSM predictions depend on the top mass 
%($\frac{\mathrm{m}^2_{\mathrm{t}}}{\mathrm{M}^2_{\mathrm{Z}}}$) 
($\mathrm{m}^2_{\mathrm{t}}/\mathrm{M}^2_{\mathrm{Z}}$) 
and to less extend to the Higgs mass 
%(log($\frac{\mathrm{M}^2_{\mathrm{H}}}{\mathrm{M}^2_{\mathrm{Z}}}$)),
(log($\mathrm{M}^2_{\mathrm{H}}/\mathrm{M}^2_{\mathrm{Z}}$)),
or to any kind of ``heavy new physics''. 

The subject of these lectures is to show how the high precision achieved in the 
EW measurements allows to test the MSM beyond the tree level 
predictions and, therefore, how this measurements are able to indirectly determine 
the value of \mt~and \MW, 
to constrain the unknown value of \MH, and at the same time to 
test the consistency between measurements and theory. At present the uncertainties 
in the theoretical predictions are dominated by the precision on the input 
parameters.

\subsection{\bf Input Parameters of the MSM}\label{subsec:input_par}

The W mass is one of the input parameters in the ``on-shell'' renormalization scheme.
It is known with a precision of about 0.04\%, although the usual procedure is 
to take \GM~(the Fermi constant measured in the muon decay) 
to predict \MW~as a function of the rest of the input parameters 
and use this more precise value. 

Therefore, the input parameters are chosen~\cite{PDG} to be: 

\noindent
\bea
 \rm{Input~Parameter}       &    \rm{Value}              & \rm{Relative~Uncertainty}       \nonumber \\
%\alpha^{-1}(0)            = & 137.0359895 (61)                             &  10^{-6} \%  \nonumber \\ 
\alpha^{-1}(\mathrm{M}^2_{\mathrm{Z}})  = & 128.933 (21)                    &  0.016    \%  \nonumber \\ 
\alpha_{\mathrm{s}}(\mathrm{M}^2_{\mathrm{Z}}) = & 0.118 (2)                &  1.1     \%  \nonumber \\ 
 G_{\mu}                  = & 1.16637 (1) \times 10^{-5}~\mathrm{GeV^{-2}}  &  0.0009  \%  \nonumber \\ 
\mathrm{M}_{\mathrm{Z}}   = & 91.1875 (21)~\mathrm{GeV}                     &  0.0023  \%  \nonumber \\
\mathrm{m}_{\mathrm{t}}   = & 174.3   (51)~\mathrm{GeV}                     &  2.9     \%  \nonumber \\
\mathrm{M}_{\mathrm{H}}   > & 114.1~\mathrm{GeV}~@95\%~C.L.                 &              \nonumber 
\eea

Notice that the less well known parameters are \mt, \als~and, of course, 
the unknown value of \MH. The next less well known parameter is 
$\alpha^{-1}(\mathrm{M}^2_{\mathrm{Z}})$, even though its value at 
$q^2 \sim 0$ is known with an amazing relative precision of $4 \times 10^{-9}$,
($\alpha^{-1}(0)=137.03599976~(50)$). 

The reason for this loss of precision when one computes the running of $\alpha$,

\noindent
\bea
 \alpha^{-1}(\mathrm{M}^2_{\mathrm{Z}}) & = & \frac{\alpha^{-1}(0)}{1-\Pi_{\gamma\gamma}}  \nonumber
\eea

\noindent is the large contribution from the light fermion loops to the photon 
vacuum polarisation, $\Pi_{\gamma\gamma}$. 
The contribution from leptons and top quark loops is well calculated~\cite{STEINHAUSSER}: 
$\Pi^{\rm{lepton}}_{\gamma\gamma} = -0.031498$ and 
$\Pi^{\rm{top}}_{\gamma\gamma} = -0.000076$ with $\mathrm{m}_{\mathrm{t}} = 174.3$~GeV.   
But for the light quarks non-perturbative
QCD corrections are large at low energy scales. The method so far
has been to use the measurement of the hadronic cross section 
through one-photon exchange, normalised to the point-like muon 
cross-section, R(s), and evaluate the dispersion integral:
 
\noindent
\bea
\Re(\Pi^{\mathrm{had}}_{\gamma\gamma}) & = & \frac{\alpha \mathrm{M}^2_{\mathrm{Z}}}{3 \pi} \Re 
\left( \int \frac{R(s')}{s'(s'-\mathrm{M}^2_{\mathrm{Z}}+i\epsilon)} ds' \right) 
\eea

\noindent giving~\cite{PIETRZYK} $\Pi^{had}_{\gamma\gamma} = - 0.02755 \pm 0.00046$, the error being 
dominated by the experimental uncertainty in the cross section measurements.

Recently, several new {\em ``theory driven''} calculations~\cite{DAVIER}~\cite{NEWALPHA} 
have reduced this error,
% ($0.016 \%$~relative precision),
by extending the regime of applicability of Perturbative QCD (PQCD). These new calculations have been 
validated by the most recent data at %$\sqrt{s} \sim$~1~to~7~GeV from 
BESS~II~\cite{BESS}.
%, included in the evaluation in reference~\cite{PIETRZYK}. 
Therefore, along these lectures, the most precise value from 
reference~\cite{DAVIER},$\Pi^{had}_{\gamma\gamma} = - 0.02763 \pm 0.00016$, will be consistently used.

%This needs to 
%be confirmed using precision measurements of the hadronic cross section at 
%$\sqrt{s} \sim$~1~to~7~GeV. The very preliminary first results from 
%BESS II~\cite{BESS} seem to validate this procedure, being in agreement with 
%the predictions from PQCD.
 
\subsection{\bf What are we measuring to test the MSM?}\label{subsec:what_meas}

From the point of view of radiative corrections we can divide the experimental measurements 
into four different groups: the \Z0~total and partial widths,
% (\GZ,\Gl,\Gh,...), 
the partial width into b-quarks ($\Gamma_{\mathrm{b}}$), the \Z0~asymmetries (\SEF) and
the W mass (\MW). 
For instance,
the leptonic width (\Gl) is mostly sensitive to isospin-breaking loop corrections 
($\Delta \rho$), the asymmetries are specially sensitive to radiative corrections to 
the \Z0~self-energy ($\Delta \kappa$ ), and \Rb~is mostly sensitive to vertex corrections ($\epsilon_b$)
in the decay 
$\mathrm{Z} \rightarrow b \bar{b}$. One more parameter,
$\Delta r$, is necessary to describe the radiative corrections to the relation 
between \GM~and \MW.

The sensitivity of these three \Z0~observables and \MW~to the input parameters is shown in 
table~\ref{tab:sensitiv}. The most sensitive observable to the unknown 
value of \MH~are the \Z0~asymmetries parametrised via \SEF. However also the sensitivity 
of the rest of the observables is very relevant compared to the achieved experimental precision.

\begin{table}[t]
\caption{ Relative error in units of per-mil on the MSM predictions 
induced by the uncertainties on the input parameters. The second column shows 
the present experimental errors.\label{tab:sensitiv}}
\vspace{0.2cm}
\begin{center}
\footnotesize
\begin{tabular}{|l|c|c c c c|}
\hline
{} &  {Exp. error} &
{$\Delta$\mt =} & {$\Delta$\MH=} &
{$\Delta$\als =} & {$\Delta\alpha^{-1}$=} \\
{} & { } &
{$\pm$5.1~GeV} & {[114-1000]~GeV} &
{$\pm$0.002} & {$\pm$0.021} \\
\hline
$\Gamma_{\mathrm{Z}}$      &  0.9  &     0.5    & \bf{3.3}  &  0.5  &   -  \\
$\mathrm{R}_{\mathrm{b}}$  &  3.0  &     0.8    &     0.1   &   -   &   -  \\
\MW                        &  0.4  & \bf{0.4}   & \bf{2.1}  &   -   &   -  \\
\SEF                       &  0.7  & \bf{0.7}   & \bf{5.2}  &   -   &  0.2 \\
\hline
\end{tabular}
\end{center}
\end{table}

\section{\bf \Z0~lineshape}

The shape of the resonance is completely characterised by three parameters: the position of the 
peak (\MZ), the width (\GZ) and the height (\S0) of the resonance:

\noindent
\bea
\sigma^0_{\mathrm{f\bar{f}}} & = & \frac{12\pi}{\mathrm{M}^2_{\mathrm{Z}}} 
         \frac{\Gamma_{\mathrm_{e}}\Gamma_{\mathrm_{f}}}{\Gamma^2_{\mathrm_{Z}}} 
\eea

The good capabilities 
of the LEP detectors to identify the lepton flavours allow to measure the ratio of 
the different lepton species with respect to the hadronic cross-section, 
\Rl = $\frac{\Gamma_{\mathrm{h}}}{\Gamma_{\mathrm{l}}}$. 

About 16~million \Z0~decays have been analysed by the four LEP collaborations, 
leading to a statistical precision on \S0 of 0.03 \% ! Therefore, 
the statistical error is not the limiting factor, but more the experimental systematic 
and theoretical uncertainties.

The error on the measurement of \MZ~is dominated by the uncertainty on the absolute 
scale of the LEP energy measurement (about 1.7~MeV), 
while in the case of \GZ~it is the point-to-point 
energy and luminosity errors which matter (about 1.3~MeV). 
The error on \S0 is 
dominated by the theoretical uncertainty on the small angle bhabha calculations 
(0.06 \%) and the uncertainty on the position of the inner edge of the luminometers
(0.07 \%). 
%but this is going to improve very soon with the new estimation 
%of this uncertainty (0.06 \%) shown at this workshop~\cite{WARD}. Moreover, a QED 
%uncertainty estimated to be around 0.05 \% has also been included in the fits.

The results of the lineshape fit are shown in table~\ref{tab:lineshape} with and 
without the hypothesis of lepton universality. From them, the leptonic widths 
and the invisible \Z0~width are derived.
 
\begin{table}[t]
\caption{Average line shape parameters from the results of the four LEP experiments.
\label{tab:lineshape}}
\vspace{0.2cm}
\begin{center}
\footnotesize
\begin{tabular}{|l|c|c|}
\hline
\raisebox{0pt}[13pt][7pt]{Parameter} &
\raisebox{0pt}[13pt][7pt]{Fitted Value} & 
\raisebox{0pt}[13pt][7pt]{Derived Parameters} \\
\hline
\MZ                       &    91187.5 $\pm$ 2.1~MeV  &  \\
\GZ                       &    2495.2  $\pm$ 2.3~MeV  &  \\
$\sigma^0_{\mathrm{had}}$ &    41.540  $\pm$ 0.037~nb &  \\
$\mathrm{R}_{\mathrm{e}}$ &    20.804  $\pm$ 0.050    & $\Gamma_{\mathrm{e}}$  = 83.92  $\pm$ 0.12~MeV \\
$\mathrm{R}_{\mu}$        &    20.785  $\pm$ 0.033    & $\Gamma_{\mu}$         = 83.99  $\pm$ 0.18~MeV \\
$\mathrm{R}_{\tau}$       &    20.764  $\pm$ 0.045    & $\Gamma_{\tau}$        = 84.08  $\pm$ 0.22~MeV \\
\hline
\multicolumn{3}{|c|}{\raisebox{0pt}[12pt][6pt]{With Lepton Universality}} \\
\hline
                          &                           & $\Gamma_{\mathrm{had}}$= 1744.4 $\pm$  2.0~MeV \\
$\mathrm{R}_{\mathrm{l}}$ &    20.767  $\pm$ 0.025    & $\Gamma_{\mathrm{l}}$  = 83.984  $\pm$ 0.086~MeV \\
                          &                           & $\Gamma_{\mathrm{inv}}$= 499.0  $\pm$  1.5~MeV \\
\hline
\end{tabular}
\end{center}
\end{table}

From the measurement of the \Z0~invisible width, and assuming the ratio of the partial widths to 
neutrinos and leptons to be the MSM predictions 
($\frac{\Gamma_{\nu}}{\Gamma_{\mathrm{l}}} = 1.9912 \pm 0.0012 $), the number of light neutrinos 
species is measured to be

\noindent
\bea
\mathrm{N}_{\nu} & = & 2.9841 \pm 0.0083. \nonumber
\eea

Alternatively, one can assume three neutrino species and determine the width from additional 
invisible decays of the \Z0~to be $\Delta\Gamma_{\mathrm{inv}}<2.0$~MeV~@95\%~C.L.

The measurement of \Rl~and $\sigma^0_{\mathrm{had}}$ are very sensitive to PQCD corrections 
and allow one of the most precise and clean determinations of \als.  
A combined fit to the measurements shown in table~\ref{tab:lineshape}, and 
imposing \mt=174.3$\pm$5.1~GeV as a constraint gives: 

\noindent
\bea
\alpha_{\mathrm{s}}(\mathrm{M}^2_{\mathrm{Z}}) & = & 0.117 \pm 0.003  \nonumber
\eea

\noindent in agreement with the world average~\cite{PDG} 
$\alpha_{\mathrm{s}}(\mathrm{M}^2_{\mathrm{Z}}) = 0.118 \pm 0.002$. 

\subsection{\bf Heavy flavour results}\label{subsec:HF}

The large mass and long lifetime of the $b$ and $c$ quarks provides a way to perform flavour tagging. 
This allows for precise measurements of the partial widths of the decays \Z0$\rightarrow c \bar{c}$ and 
\Z0$\rightarrow b \bar{b}$. It is useful to normalise the partial width to \Gh~by 
measuring the partial decay fractions with respect to all hadronic decays

\noindent
\bea
\mathrm{R}_{\mathrm{c}} \equiv \frac{\Gamma_{c}}{\Gamma_{\mathrm{h}}} & , & 
\mathrm{R}_{\mathrm{b}} \equiv \frac{\Gamma_{b}}{\Gamma_{\mathrm{h}}}.  \nonumber 
\eea

With this definition most of the radiative corrections appear both in the numerator 
and denominator and thus cancel out, with the important exception of the vertex corrections 
in the \Z0 $b\bar{b}$ vertex. This is the only relevant correction to \Rb, and within the 
MSM basically depends on a single parameter, the mass of the top quark.

The partial decay fractions of the \Z0~to other quark flavours, like \Rc, are only weakly 
dependent on \mt; the residual weak dependence is indeed due to the presence of 
$\Gamma_{b}$ in the denominator. The MSM predicts \Rc = 0.172, valid over a wide 
range of the input parameters.

The combined values from the measurements of LEP and SLD gives

\noindent
\bea
\mathrm{R}_{\mathrm{b}} & = & 0.21646 \pm 0.00065 \nonumber \\
\mathrm{R}_{\mathrm{c}} & = & 0.1719  \pm 0.0031  \nonumber 
\eea

\noindent with a correlation of -14\% between the two values. 
%The large sensitivity 
%of \Rb~to the top mass allows to determine indirectly its mass to be \mt=151$\pm$25~GeV, 
%in agreement with the direct measurement (\mt=173.8$\pm$5.0~GeV).

\section{\bf \Z0~asymmetries: $\boldmath{\sin^2\theta_{\mathrm{eff}}}$ }

Parity violation in the weak neutral current is caused by the difference of couplings 
of the \Z0~to right-handed and left-handed fermions. If we define $A_{\mathrm{f}}$ as

\noindent
\bea
A_{\mathrm{f}} & \equiv & 
\frac{2 \biggl( \frac{g^f_V}{g^f_A} \biggr) }{1 + \biggl( \frac{g^f_V}{g^f_A}\biggr)^2},
\eea

\noindent where $g^f_{V(A)}$ denotes the vector(axial-vector) coupling constants, one can write 
all the \Z0~asymmetries in terms of $A_{\mathrm{f}}$.

Each process $e^+ e^- \rightarrow \mathrm{Z}^{0} \rightarrow \mathrm{f}\bar{\mathrm{f}}$ 
can be characterised 
by the direction and the helicity of the emitted fermion (f). Calling forward the hemisphere 
into which the electron beam is pointing, the events can be subdivided into four categories: 
FR,BR,FL and BL, corresponding to right-handed (R) or left-handed (L) fermions emitted 
in the forward (F) or backward (B) direction. Then, one can write three \Z0~asymmetries 
as:

\noindent
\bea
A_{\mathrm{pol}} \equiv & 
\frac{\sigma_{\mathrm{FR}}+\sigma_{\mathrm{BR}}-\sigma_{\mathrm{FL}}-\sigma_{\mathrm{BL}}}
     {\sigma_{\mathrm{FR}}+\sigma_{\mathrm{BR}}+\sigma_{\mathrm{FL}}+\sigma_{\mathrm{BL}}}
                        & = -A_{\mathrm{f}} \\
A^{\mathrm{FB}}_{\mathrm{pol}} \equiv & 
\frac{\sigma_{\mathrm{FR}}+\sigma_{\mathrm{BL}}-\sigma_{\mathrm{BR}}-\sigma_{\mathrm{FL}}}
     {\sigma_{\mathrm{FR}}+\sigma_{\mathrm{BR}}+\sigma_{\mathrm{FL}}+\sigma_{\mathrm{BL}}}
                        & = -\frac{3}{4} A_{\mathrm{e}}  \\
A_{\mathrm{FB}} \equiv & 
\frac{\sigma_{\mathrm{FR}}+\sigma_{\mathrm{FL}}-\sigma_{\mathrm{BR}}-\sigma_{\mathrm{BL}}}
     {\sigma_{\mathrm{FR}}+\sigma_{\mathrm{BR}}+\sigma_{\mathrm{FL}}+\sigma_{\mathrm{BL}}}
                        & = \frac{3}{4} A_{\mathrm{e}}A_{\mathrm{f}} 
\eea

\noindent and in case the initial state is polarised with some degree of polarisation 
($P$), one can define:

\noindent
\bea
A_{\mathrm{LR}} \equiv & \frac{1}{P} 
\frac{\sigma_{\mathrm{Fl}}+\sigma_{\mathrm{Bl}}-\sigma_{\mathrm{Fr}}-\sigma_{\mathrm{Br}}}
     {\sigma_{\mathrm{Fr}}+\sigma_{\mathrm{Br}}+\sigma_{\mathrm{Fl}}+\sigma_{\mathrm{Bl}}}
                        & = A_{\mathrm{e}} \\
A^{\mathrm{pol}}_{\mathrm{FB}} \equiv & -\frac{1}{P} 
\frac{\sigma_{\mathrm{Fr}}+\sigma_{\mathrm{Bl}}-\sigma_{\mathrm{Fl}}-\sigma_{\mathrm{Br}}}
     {\sigma_{\mathrm{Fr}}+\sigma_{\mathrm{Br}}+\sigma_{\mathrm{Fl}}+\sigma_{\mathrm{Bl}}}
                        & = \frac{3}{4} A_{\mathrm{f}} 
\eea

\noindent where r(l) denotes the right(left)-handed initial state polarisation. Assuming lepton 
universality, all this observables depend only on the ratio between the vector and axial-vector 
couplings. It is conventional to define the effective mixing angle \SEF~as

\noindent
\bea
\sin^2\theta_{\mathrm{eff}} & \equiv & \frac{1}{4} \biggl( 1 - \frac{g^l_V}{g^l_A}  \biggr)
\eea

\noindent and to collapse all the asymmetries into a single parameter \SEF.

\subsection{\bf Lepton asymmetries}\label{subsec:lepton_asym}
\subsubsection{\bf Angular distribution}\label{subsec:lepton_afb}

The lepton forward-backward asymmetry is measured from the angular distribution 
of the final state lepton.
The measurement of $A^{\mathrm{l}}_{FB}$ is quite simple and robust and its 
accuracy is limited by the statistical error. The common systematic uncertainty in the LEP measurement 
due to the uncertainty on the LEP energy measurement is about~0.0003.
The values measured by the LEP collaborations are in agreement with lepton universality,

\noindent
\bea
 & A^e_{\mathrm{FB}}     = 0.0145 \pm 0.0025  &    \nonumber \\
 & A^{\mu}_{\mathrm{FB}} = 0.0169 \pm 0.0013  &    \nonumber \\
 & A^{\tau}_{\mathrm{FB}}= 0.0188 \pm 0.0017  &    \nonumber 
\eea

\noindent and can be combined into a single measurement of \SEF,

\noindent
\bea
A^{\mathrm{l}}_{\mathrm{FB}}= 0.01714 \pm 0.00095 & \Longrightarrow &
{ \sin^2\theta_{\mathrm{eff}} = 0.23099 \pm 0.00053}. \nonumber 
\eea

\subsubsection{\bf Tau polarisation at LEP}\label{subsec:lepton_taupol}

Tau leptons decaying inside the apparatus acceptance can be used to measure 
the polarised asymmetries defined by equations~(4) and~(5). A more sensitive method 
is to fit the measured dependence of $A_{\mathrm{pol}}$ as a function of the 
polar angle $\theta$ :

\noindent
\bea
A_{\mathrm{pol}}(\cos\theta) & =  & 
- \frac{A_{\tau}(1+\cos^2\theta)+2 A_e \cos\theta}{(1+\cos^2\theta) + 2 A_{\tau} A_e \cos\theta }
\eea

The sensitivity of this measurement to \SEF~is larger because the dependence 
on $A_{\mathrm{l}}$ is linear to a good approximation. 
The accuracy of the measurements is dominated by the statistical 
error. The typical systematic error is about 0.003 for $A_{\tau}$ and 0.001 for $A_e$.
The LEP measurements are:

\noindent
\bea
A_{e}   = 0.1498 \pm 0.0049 & \Longrightarrow &
{ \sin^2\theta_{\mathrm{eff}} = 0.23117 \pm 0.00062} \nonumber  \\
A_{\tau}= 0.1439 \pm 0.0043 & \Longrightarrow &
{ \sin^2\theta_{\mathrm{eff}} = 0.23192 \pm 0.00054} \nonumber  
\eea

\subsection{\bf \ALR~from SLD}\label{subsec:alr}

The linear accelerator at SLAC (SLC) allows to collide positrons 
with a highly longitudinally polarised electron beam (up to 77\% polarisation). Therefore, 
the SLD detector can measure the left-right cross-section asymmetry (\ALR) 
defined by equation~(7). This observable is a factor of 4.6 times more sensitive to 
\SEF~than, for instance, $A^{\mathrm{l}}_{\mathrm{FB}}$ for a given precision.
The measurement is potentially free of experimental 
systematic errors, with the exception of the polarisation measurement that 
has been carefully cross-checked at the 1\% level. SLD final  
measurement gives

\noindent
\bea
A_{\mathrm{LR}}   = 0.1514 \pm 0.0022 & \Longrightarrow &
{ \sin^2\theta_{\mathrm{eff}} = 0.23097 \pm 0.00027}, \nonumber 
\eea

\noindent and assuming lepton universality it can be 
combined with measurements at SLD of the lepton
left-right forward-backward asymmetry ($A^{\mathrm{pol}}_{\mathrm{FB}}$) 
defined in equation~(8) to give

\noindent
\bea
         & {\sin^2\theta_{\mathrm{eff}} = 0.23098 \pm 0.00026}. &  \nonumber 
\eea

\subsection{\bf Lepton couplings}\label{subsec:lepton_coup}

All the previous measurements of the lepton coupling ($A_{\mathrm{l}}$) can be 
combined with a $\chi^2/\mathrm{dof}=2.6/3$ and give
\noindent
\bea
A_{\mathrm{l}}   = 0.1501 \pm 0.0016 & \Longrightarrow &
{\sin^2\theta_{\mathrm{eff}} = 0.23113 \pm 0.00021}. \nonumber 
\eea
 
\begin{figure}
\centering
\mbox{%
\epsfig{file=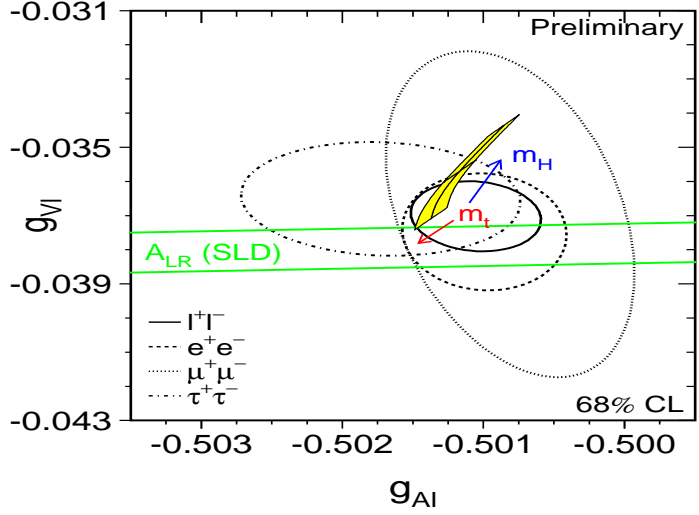 % this is the name of your figure file
        ,height=8cm  % this is the height of the figure (optional)
        ,width=10cm   % this is the width of the figure (optional)
       }%
}
\caption{Contours of 68\% probability in the $g^l_V - g^l_A$ plane. The solid 
contour results from a fit assuming lepton universality. Also shown is the one 
standard deviation band resulting from the \ALR~measurement of SLD.} % the figure caption
\label{fig:gvga}      % the (optional) label to refer to in the text
\end{figure}

The asymmetries measured are only sensitive to the ratio between the 
vector and axial-vector couplings. If we introduce also the measurement 
of the leptonic width shown in table~\ref{tab:lineshape} we can 
fit the lepton couplings to the \Z0~to be

\noindent
\bea
g^l_V & = & -0.03783 \pm 0.00041, \nonumber \\
g^l_A & = & -0.50123 \pm 0.00026, \nonumber 
\eea

\noindent where the sign is chosen to be negative by definition. 
Figure~\ref{fig:gvga} shows the 68~\% probability contours in the 
$g^l_V - g^l_A$ plane. 

%Notice that the fitted value of $g^l_A$ is 
%different from the Born prediction ($-\frac{1}{2}$) by about $3.4 \sigma$, 
%showing evidence for radiative corrections in the $\rho$ parameter.

\subsection{\bf Quark asymmetries}\label{subsec:quark_asym}

\subsubsection{\bf Heavy Flavour asymmetries}\label{subsec:HF_asym}

The inclusive measurement of the $b$ and $c$ asymmetries is more sensitive 
to \SEF~than, for instance, the leptonic forward-backward asymmetry. The 
reason is that $A_b$ and $A_c$ are mostly independent of \SEF, therefore 
$A^{b(c)}_{\mathrm{FB}}$ (which is proportional to the product $A_e A_{b(c)}$)
is a factor 3.3(2.4) more sensitive than $A^{\mathrm{l}}_{\mathrm{FB}}$.
The typical systematic uncertainty in $A^{b(c)}_{\mathrm{FB}}$ is 
about 0.001(0.002) and the precision of the measurement is  
dominated by statistics. 

SLD can measure also the $b$ and $c$
left-right forward-backward asymmetry defined in equation~(8) which is 
a direct measurement of the quark coupling $A_b$ and $A_c$. 
The combined fit for the LEP and SLD measurements gives

\noindent
\bea
 A^b_{\mathrm{FB}}= 0.0990 \pm 0.0017  & \Longrightarrow & {\sin^2\theta_{\mathrm{eff}} = 0.23226 \pm 0.00031} 
\nonumber \\
 A^c_{\mathrm{FB}}= 0.0685 \pm 0.0034  & \Longrightarrow & {\sin^2\theta_{\mathrm{eff}} = 0.23272 \pm 0.00079} 
\nonumber \\
 A_b              = 0.922  \pm 0.020   &   &  \nonumber \\
 A_c              = 0.670  \pm 0.026   &   &  \nonumber 
\eea

\noindent where 15\% is the largest correlation between $A^b_{\mathrm{FB}}$ and $A^c_{\mathrm{FB}}$.

Taking the value of $A_{\mathrm{l}}$ given in section~\ref{subsec:lepton_coup} and these 
measurements together in a combined fit gives

\noindent
\bea
 & A_b              = 0.879  \pm 0.018   &    \nonumber \\
 & A_c              = 0.608  \pm 0.031   &    \nonumber 
\eea

\noindent to be compared with the MSM predictions $A_b = 0.935$ and 
$A_c = 0.668$ valid over a wide 
range of the input parameters.
%where the uncertainty in the predictions due to the input parameters is negligible. 
The measurement of $A_c$ is 1.9 standard deviations lower than the predicted value,
while the measurement of $A_b$ is 3.1 standard deviations lower. 
Notice that the direct measurements of $A_{b(c)}$ at SLD 
are in perfect agreement with the MSM prediction. The deviation in the combined fit of $A_{b(c)}$ and $A_l$, 
is just a reflection of the discrepancy between leptonic and hadronic measurements of 
$\sin^2\theta_{\mathrm{eff}}$, (see section~\ref{subsec:sineff}).
 
%This is due to three independent measurements: the SLD measurement of 
%$A_b$ is low compared with the MSM, while the LEP measurement of 
%$A^b_{\mathrm{FB}}$ is low and the SLD measurement of 
%$A_{\mathrm{LR}}$ is high compared with the results of the best fit to 
%the MSM predictions (see section~\ref{subsec:fit}).

%This is due to three independent measurements: the 
%LEP measurement of $A^b_{\mathrm{FB}}$ is low compared with the MSM, 
%the SLD measurement of $A_b$ is also low, and the SLD measurement of 
%$A_{\mathrm{LR}}$ is high compared with the MSM.

\subsubsection{\bf Jet charge asymmetries}\label{subsec:jet_asym}

The average charge flow in the inclusive samples of hadronic \Z0~decays 
is related to the forward-backward asymmetries of individual quarks:

\noindent
\bea
 \langle \mathrm{Q}_{\mathrm{FB}} \rangle & = & \sum_{\mathrm{q}} 
\delta_{\mathrm{q}} A^{\mathrm{q}}_{\mathrm{FB}} \frac{\Gamma_{\mathrm{q\bar{q}}}}{\Gamma_{\mathrm{h}}}   
\eea

\noindent where $\delta_{\mathrm{q}}$, the charge separation, is the average charge 
difference between the quark and antiquark hemispheres in an event.
The combined LEP value is

\noindent
\bea
         & {\sin^2\theta_{\mathrm{eff}} = 0.2324 \pm 0.0012}. &  \nonumber 
\eea

\subsection{\bf Comparison of the determinations of 
$\boldmath{\sin^2\theta_{\mathrm{eff}}}$}\label{subsec:sineff}

The combination of all the quark asymmetries shown in this section can be 
directly compared to the determination of \SEF~obtained with leptons, 

\noindent
\bea
{\sin^2\theta_{\mathrm{eff}} = 0.23230 \pm 0.00029} &   & \mathrm{(quark-asymmetries)} \nonumber \\ 
{\sin^2\theta_{\mathrm{eff}} = 0.23113 \pm 0.00021} &   & \mathrm{(lepton-asymmetries)} \nonumber 
\eea

\noindent which shows a 3.3~$\sigma$ discrepancy.

\begin{figure}
\centering
\mbox{%
\epsfig{file=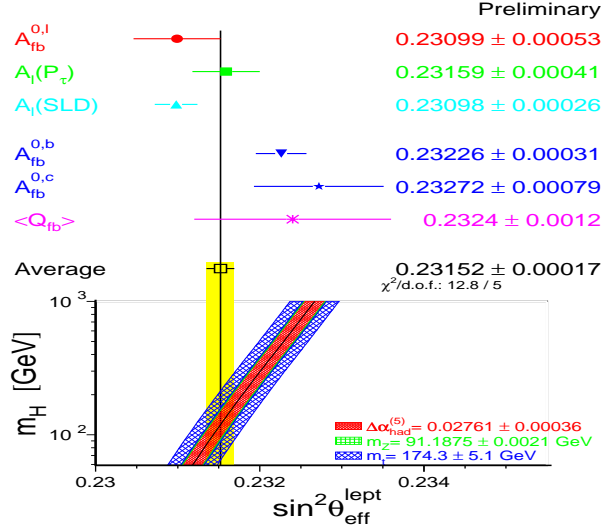 % this is the name of your figure file
        ,height=7cm  % this is the height of the figure (optional)
        ,width=8cm   % this is the width of the figure (optional)
       }%
}
\caption{Comparison of several determinations of \SEF~from asymmetries.} % the figure caption
\label{fig:sef}      % the (optional) label to refer to in the text
\end{figure}

Over all, the agreement is acceptable, and the combination of the individual 
determinations of \SEF~gives

\noindent
\bea
         & {\sin^2\theta_{\mathrm{eff}} = 0.23152 \pm 0.00017} &  \nonumber 
\eea

\noindent with a $\chi^2/\mathrm{dof}=12.8/5$ as it is shown 
in figure~\ref{fig:sef}.

\section{W mass}

Since 1996 up to 2000 LEP has been running at energies above the W-pair 
production treshold and about 40k W-pairs have been detected by the LEP experiments. 
The cross-section for the process $e^+ e^- \rightarrow W^+ W^-$ 
has been measured with a precision of $1 \%$. The theoretical calculations 
have been updated to match with this precision, confirming the indirect evidence 
from \Z0 physics of Gauge Boson Couplings predicted by the MSM.

More interesting in the context of these lectures, is the improvement on 
the W mass accuracy, previously measured in proton-proton or $\nu$N collisions.

\subsection{\bf  W mass at pp colliders }

At hadron colliders, the W mass is obtained from the distribution of the W transverse 
mass, that is the invariant mass of the W decay products evaluated in the plane transverse 
to the beam. This is because the longitudinal component of the neutrino momentum cannot 
be measured in a pp collider. On the other hand, the transverse momentum of the neutrino 
can be deduced from the vector sum of the transverse momentum of the charged lepton and the 
transverse momentum of the system recoiling against the W.

The uncertainty on the W mass is dominated by the uncertainty in the lepton energy/momentum 
calibration. The combination of the measurements at FERMILAB (CDF/D0), and CERN (UA2) gives:  

\noindent
\bea
         & {\mathrm{M}_{\mathrm{W}} = 80.454 \pm 0.060~\mathrm{GeV}} &  \nonumber 
\eea

\noindent where the error is dominated by the systematic uncertainty (50 MeV).

\subsection{\bf W mass from $\nu$N scattering}

The ratio between the neutral and charged current interaction of neutrino on 
isoscalar targets, provides an indirect determination of the ratio between 
the W mass and the \Z0 mass. The main experimental uncertainty comes from the 
model needed for the background subtraction. The results from NuTeV combined with 
previous CCFR measurements (asuming \mt=175~GeV and \MH=150~GeV) gives:

\noindent
\bea
         & {\mathrm{M}_{\mathrm{W}} = 80.25 \pm 0.11~\mathrm{GeV}} &  \nonumber 
\eea

\noindent where the dependence with \mt~and \MH~has to be taken into account 
in the MSM fits in section~\ref{subsec:fit}.

\subsection{\bf W mass at LEP}

The W-pair production cross-section near the treshold has a strong dependence 
on the W mass. The first data collected at LEP just above treshold has been used 
to get a measurement of the W mass:

\noindent
\bea
         & {\mathrm{M}_{\mathrm{W}} = 80.40 \pm 0.22~\mathrm{GeV}} &  \nonumber 
\eea

But the most precise measurement of the W mass comes from the kinematic reconstruction 
of the W decay products at LEP. The precise knowledge of the c.o.m. energy is used 
to improve the experimental resolutions. The W mass is extracted from a comparison 
between data and Monte Carlo simulation for different values of the W mass giving:

\noindent
\bea
         & {\mathrm{M}_{\mathrm{W}} = 80.450 \pm 0.039~\mathrm{GeV}} &  \nonumber 
\eea

The measurement is dominated by systematic uncertainties (30 MeV). The main systematic
uncertainty is due to the hadronization model (17~MeV) and to the knowledge of the 
LEP c.o.m. energy (17~MeV) which affects both channels in a coherent way: 4q channel where 
both W's decay into quarks, and 2q channel when one of the W's decay into a lepton and a neutrino.

There are other systematic sources related to the hadronization model that only affect the
4q channel.
In particular, 
the separation of the decay vertices is about 0.1~fm, which is small compared with the typical 
hadronization scale of 1~fm. This fact may lead to non-perturbative phenomena 
interconnecting the decays of the two W's and introducing a source of systematic 
uncertainties in the measurement. By comparing diferent models of Colour Reconnection (CR) 
that reproduce the precise LEP data, a maximum shift of 40~MeV is quoted in the 
4q channel. A similar procedure is used to quote a maximum shift of 25~MeV due to 
possible Bose-Einstein (BE) correlations between decay products from the two W's.

As this phenomena only affects the W mass obtained from the 4q channel at LEP, it is 
instructive to check the consistency of this measurement with respect to the rest of the 
W mass measurements: 

\noindent
\bea
& {\mathrm{M}_{\mathrm{W}}(4q~LEP                     ) = 80.457 \pm 0.030(stat.) \pm 0.026(syst.) \pm 0.047(CR/BE)~\mathrm{GeV}} &  \nonumber \\
& {\mathrm{M}_{\mathrm{W}}(2q~LEP~+~CDF/D0~+~xsect~LEP) = 80.449 \pm 0.024(stat.) \pm 0.025(syst.)~\mathrm{GeV}} &  \nonumber
\eea

\noindent which leads to a non-significative difference of $+8 \pm 39$~MeV, and supports 
that the quoted systematic uncertainty for these effects (47~MeV for the 4q channel) 
is reasonable. 

All direct measurements of the W mass from LEP and TEVATRON can be combined 
to give a world average value of:

\noindent
\bea
         & {\mathrm{M}_{\mathrm{W}} = 80.451 \pm 0.033~\mathrm{GeV}} &  \nonumber 
\eea

\section{Consistency with the Minimal Standard Model}\label{sec:consistency}

The MSM predictions are computed using the programs TOPAZ0~\cite{TOPAZ0} 
and ZFITTER~\cite{ZFITTER}. They represent the state-of-the-art in the 
computation of radiative corrections, and incorporate recent calculations such as
the QED radiator function to \cal{O}($\alpha^3$), four-loop 
QCD effects, non-factorisable QCD-EW corrections, 
and two-loop sub-leading 
\cal{O}($\alpha^2 \mathrm{m}^2_{\mathrm{t}} / \mathrm{M}^2_{\mathrm{Z}}$) 
corrections, resulting in a significantly reduced theoretical 
uncertainty compared to the work summarized in reference~\cite{YR95}.

\subsection{\bf Are we sensitive to radiative corrections other than $\Delta\alpha$?}\label{subsec:sensitivity}

This is the most natural question to ask if one pretends to test the MSM as 
a Quantum Field Theory and
to extract information on the only unknown parameter in the MSM, \MH.
%In fact, the answer is already given in table~\ref{tab:sensitiv}, 
%where one can see the different sensitivity of the MSM prediction on the 
%input parameters. 

The MSM prediction of \Rb~neglecting radiative corrections is ${\mathrm{R}}^0_{\mathrm{b}}=0.2183$, 
while the measured value given in section~\ref{subsec:HF} is about $2.8\sigma$ lower.
From table~\ref{tab:sensitiv} one can see that the MSM prediction depends only 
on \mt~and allows to determine indirectly its mass to be \mt=155$\pm$20~GeV, 
in agreement with the direct measurement (\mt=174.3$\pm$5.1~GeV).

%The large sensitivity 
%of \Rb~to the top mass allows to determine indirectly its mass to be \mt=151$\pm$25~GeV, 
%in agreement with the direct measurement (\mt=173.8$\pm$5.0~GeV).

From the measurement of the leptonic width, the vector-axial coupling given in 
section~\ref{subsec:lepton_coup} disagrees with the Born prediction (-1/2) by about 
$4.7\sigma$, showing evidence for radiative corrections in the 
$\rho$ parameter, $\Delta\rho = 0.005 \pm 0.001$.

%Notice that the fitted value of $g^l_A$ is 
%different from the Born prediction ($-\frac{1}{2}$) by about $3.4 \sigma$, 
%showing evidence for radiative corrections in the $\rho$ parameter.

However, the most striking evidence for pure weak radiative corrections is not coming 
from \Z0~physics, but from \MW~and its relation with \GM. The value measured 
at LEP and TEVATRON is \MW=$80.451 \pm 0.033$~GeV. From this measurement 
and through the relation

\noindent
\bea
 \mathrm{M}^2_{\mathrm{W}} \left(1 - \frac{\mathrm{M}^2_{\mathrm{W}}}{\mathrm{M}^2_{\mathrm{Z}}} \right) & 
 = & \frac{\pi \alpha}{G_{\mu} \sqrt{2}} \left( 1 + \Delta r\right)  
\eea

\noindent one gets a measurement of $\Delta r = 0.032 \pm 0.002$, and subtracting the 
value of $\Delta\alpha$ ($\Delta\alpha = -\Pi_{\gamma\gamma}$),
given in section~\ref{subsec:input_par}, 
one obtains $\Delta r_{\mathrm{W}} = \Delta r - \Delta \alpha = -0.027 \pm 0.002$, 
which is about $12.9\sigma$ different from zero. 
%A more detailed investigation 
%on the evidence for pure weak radiative corrections can be found in reference~\cite{SIRLIN}.

\subsection{\bf Fit to the MSM predictions}\label{subsec:fit}

\begin{figure}
\centering
\mbox{%
\epsfig{file=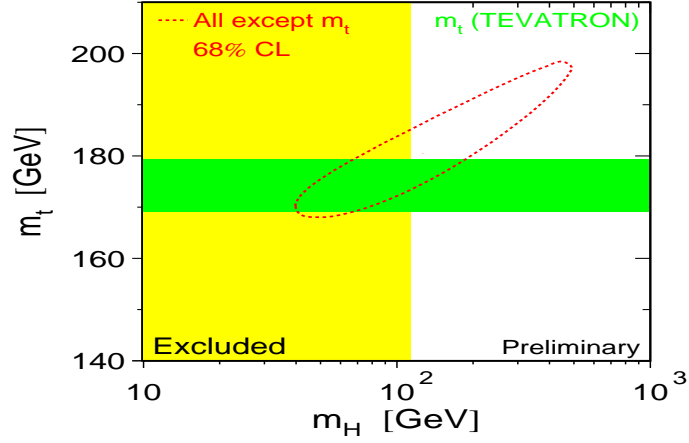 % this is the name of your figure file
        ,height=7cm  % this is the height of the figure (optional)
        ,width=9cm   % this is the width of the figure (optional)
       }%
}
\caption{The 68\% confidence level contour in the \mt~vs~\MH~plane. The 
vertical band shows the 95\% C.L. exclusion limit on \MH~from direct searches.} % the figure caption
\label{fig:mhmt}      % the (optional) label to refer to in the text
\end{figure}

\begin{figure}
\centering
\mbox{%
\epsfig{file=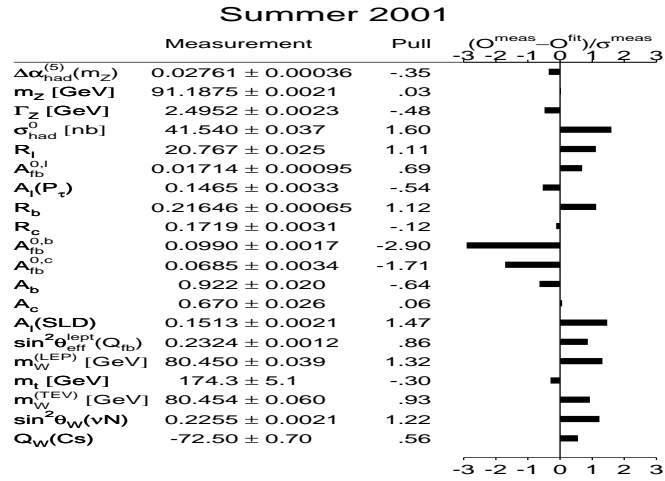 % this is the name of your figure file
        ,height=7cm  % this is the height of the figure (optional)
        ,width=9cm   % this is the width of the figure (optional)
       }%
}
\caption{Pulls of the measurements with respect to the best fit results. The 
pull is defined as the difference of the measurement to the fit prediction divided 
by the measurement error.} % the figure caption
\label{fig:pulls}      % the (optional) label to refer to in the text
\end{figure}

Having shown that there is sensitivity to pure weak corrections with the accuracy in the 
measurements achieved so far, one can envisage to fit the values of the unknown Higgs mass and 
the less well known top mass in the context of the MSM predictions. 
The fit is done using the \Z0~measurements, the W mass measurements and $\nu$N scattering 
measurements.
The quality of the fit is acceptable, ($\chi^2/\mathrm{dof}=22.7/14$) and 
the indirect determination of the top mass gives,

\noindent
\bea
\mathrm{m}_{\mathrm{t}}      & = & 180^{+11}_{-9}~\mathrm{GeV}                               \nonumber 
%\mathrm{m}_{\mathrm{t}}       = & 161.1^{+8.2}_{-7.1} \mathrm{GeV} &                              \nonumber \\ 
%\log(\mathrm{M}_{\mathrm{H}}) = & 1.51^{+0.38}_{-0.29}             & (32^{+46}_{-16} \mathrm{GeV}) \nonumber \\ 
%\alpha_s                      = & 0.120 \pm 0.003                  &  \nonumber
\eea

\noindent to be compared with \mt=174.3$\pm$5.1~GeV measured at TEVATRON. The result 
of the fit is shown in the \MH-\mt~plane in figure~\ref{fig:mhmt}. Both 
determinations of \mt~have similar precision and are compatible. Therefore, 
one can constrain the previous fit with the direct measurement of \mt~and obtains:

\noindent
\bea
\mathrm{m}_{\mathrm{t}}       = & 175.7 \pm 4.4~\mathrm{GeV} &                              \nonumber \\ 
\log(\mathrm{M}_{\mathrm{H}}/\mathrm{GeV}) = & 1.96 \pm 0.19       & 
                                   (\mathrm{M}_{\mathrm{H}} = 91^{+49}_{-33}~\mathrm{GeV}) \nonumber \\ 
\alpha_s                      = & 0.118 \pm 0.003            &  \nonumber
\eea

\noindent with a $\chi^2/\mathrm{dof}=23.0/15$. 
Most of the contribution to the $\chi^2$ is from the discrepancy in the hadronic 
measurements of \SEF~mentioned in section~\ref{subsec:sineff}. The distribution of the 
pulls of each measurement is shown in figure~\ref{fig:pulls}.
%The agreement 
%of the fit with the measurements is impressive and it is shown as a pull distribution in 
%figure~\ref{fig:pulls}. 

The best indirect determination of the W mass is obtained from 
the MSM fit when no information from the direct measurement is used, 

\noindent
\bea
\mathrm{M}_{\mathrm{W}}     &  =  & 80.375 \pm 0.022~\mathrm{GeV}. \nonumber
\eea

\noindent which is a bit low (-1.9$\sigma$) compared with the direct measurement at LEP and 
TEVATRON, $\mathrm{M}_{\mathrm{W}} =  80.451 \pm 0.033~\mathrm{GeV}$. As it would become 
more clear in section~\ref{subsec:consistency},  this is again a reflection of the 
discrepancy in the hadronic measurements of \SEF. The indirect prediction of the W mass 
not using  $A^q_{\mathrm{FB}}$ gives: $\mathrm{M}_{\mathrm{W}} =  80.414 \pm 0.024~\mathrm{GeV}$,
in good agreement with the direct determination of the W mass.

%Therefore 
%the aim for LEP2 and TEVATRON is to measure the W mass with at least a precision 
%of 30~MeV.

%The most significant correlation on the fitted parameters is 77\% between 
%$\log(\mathrm{M}_{\mathrm{H}}/\mathrm{GeV})$ and $\alpha(\mathrm{M}^2_{\mathrm{Z}})$. 
%If one of the more precise new evaluations of $\Delta\alpha$ mentioned in 
%section~\ref{subsec:input_par} is used, this correlation decreases dramatically 
%and the precision on $\log(\mathrm{M}_{\mathrm{H}}/\mathrm{GeV})$ improves by about 30\%. 
%For instance, using $\alpha^{-1}(\mathrm{M}^2_{\mathrm{Z}}) = 128.923 \pm 0.036$ 
%from reference~\cite{DAVIER}, one gets: 

%\noindent
%\bea
%\mathrm{m}_{\mathrm{t}}       = & 171.4 \pm 4.8~\mathrm{GeV}       &                              \nonumber \\ 
%\log(\mathrm{M}_{\mathrm{H}}/\mathrm{GeV}) = & 1.96^{+0.23}_{-0.26}       & 
%                                   (\mathrm{M}_{\mathrm{H}} = 91^{+64}_{-41}~\mathrm{GeV}) \nonumber \\ 
%\alpha_s                      = & 0.119 \pm 0.003                  &  \nonumber
%\eea

%\noindent with the same confidence level ($\chi^2/\mathrm{dof}=14.9/15$) and 
%a correlation of 39\% between 
%$\log(\mathrm{M}_{\mathrm{H}}/\mathrm{GeV})$ and $\alpha(\mathrm{M}^2_{\mathrm{Z}})$.

\section{Constraints on \MH}

In the previous section it has been shown that the global MSM fit to the data
gives
\noindent
\bea
\log(\mathrm{M}_{\mathrm{H}}/\mathrm{GeV}) = & 1.96 \pm 0.19       & 
                                   (\mathrm{M}_{\mathrm{H}} = 91^{+49}_{-33}~\mathrm{GeV}) \nonumber 
\eea

\noindent and taking into account the theoretical uncertainties (about 
$\pm$0.05 in 
$\log(\mathrm{M}_{\mathrm{H}}/\mathrm{GeV})$), 
this implies a one-sided 95\%~C.L. limit of:

\noindent
\bea
\mathrm{M}_{\mathrm{H}} &  < & 196~\mathrm{GeV}~@ 95 \%~\mathrm{C.L.}  \nonumber 
\eea

\noindent which does not take into account the limits from direct searches 
($\mathrm{M}_{\mathrm{H}} > 114.1~\mathrm{GeV}~@ 95 \%~\mathrm{C.L.}$).

\subsection{\bf Consistency of the Higgs mass determination}\label{subsec:consistency}

As described in section~\ref{subsec:what_meas}, one can divide the measurements sensitive 
to the Higgs mass into three different groups: Asymmetries ($\Delta \kappa$), Widths ($\Delta \rho$) 
and the W mass ($\Delta r$). They test 
conceptually different components of the radiative corrections and it is interesting to 
check the internal consistency. Given the discrepancies between hadronic and leptonic measurements 
of the \Z0~asymmetries, it is instructive to quote separate results for the asymmetries.

\begin{figure}
\centering
\mbox{%
\epsfig{file=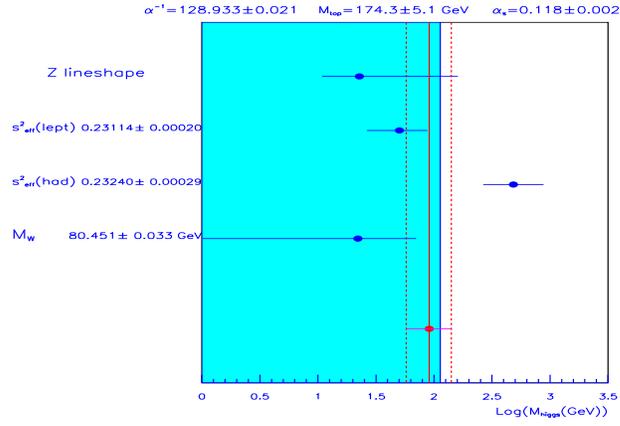 % this is the name of your figure file
        ,height=6cm  % this is the height of the figure (optional)
        ,width=9cm   % this is the width of the figure (optional)
       }%
}
\caption{Individual determination of $\log(\mathrm{M}_{\mathrm{H}}/\mathrm{GeV})$ for each 
of the measurements. The vertical band shows the 95\% C.L. exclusion limit on 
\MH~from direct searches.} % the figure caption
\label{fig:logmh}      % the (optional) label to refer to in the text
\end{figure}

Repeating the MSM fit shown in the previous section for the three different groups of measurements
with the additional constraint from~\cite{PDG} $\alpha_s = 0.118 \pm 0.002$ gives the results shown in
the second column in 
table~\ref{tab:consistency}. 
The indirect determination of \MH~from the \Z0~lineshape, from the leptonic asymmetries and from 
the W mass are 
in amazing agreement, and prefer a very low value of the Higgs mass. Only the hadronic asymmetries, 
somehow, contradict this tendency.
This is seen with more 
detail in figure~\ref{fig:logmh}, where the individual determinations of 
$\log(\mathrm{M}_{\mathrm{H}}/\mathrm{GeV})$ are shown for each measurement.

From table~\ref{tab:consistency} it is clear that any future improvement on the 
indirect determination of the Higgs mass 
needs a more precise determination of the Top mass.

\begin{table}[t]
\caption{ Results on $\log(\mathrm{M}_{\mathrm{H}}/\mathrm{GeV})$ for different samples of 
measurements. In the fit the input parameters and their uncertainties are taken to 
be the values presented in section~\ref{subsec:input_par}. The impact of the uncertainty in each 
parameter is explicitely shown. \label{tab:consistency}}
\vspace{0.2cm}
\begin{center}
\footnotesize
\begin{tabular}{|l|c|c c c|}
\hline
{} & \raisebox{0pt}[13pt][7pt]{$\log(\mathrm{M}_{\mathrm{H}})$} &
\raisebox{0pt}[13pt][7pt]{$[\Delta\log(\mathrm{M}_{\mathrm{H}})]^2$}  & = &   
\raisebox{0pt}[13pt][7pt]{$[\Delta_{\mathrm{exp.}}]^2 $}        + 
\raisebox{0pt}[13pt][7pt]{$[\Delta\mathrm{m}_{\mathrm{t}}]^2 $} + 
\raisebox{0pt}[13pt][7pt]{$[\Delta\alpha]^2 $}                  + 
\raisebox{0pt}[13pt][7pt]{$[\Delta\alpha_s]^2$} \\
\hline
\raisebox{0pt}[13pt][7pt]{Had. Asymm.}                                    & $2.68 \pm 0.26$ & 
$[0.26]^2$ & = & $[0.22]^2$ +  $[0.14]^2$ +  $[0.04]^2$ +  $[0.01]^2$     \\ 
\raisebox{0pt}[13pt][7pt]{Lep. Asymm.}                                    & $1.70 \pm 0.26$ & 
$[0.26]^2$ & = & $[0.22]^2$ +  $[0.14]^2$ +  $[0.04]^2$ +  $[0.01]^2$     \\ 
\raisebox{0pt}[10pt][7pt]{\Z0 lineshape}                                         & $1.36^{+0.85}_{-0.32}$ &  
$[0.46]^2$ & = & $[0.43]^2$ +  $[0.14]^2$ +  $[0.02]^2$ +  $[0.08]^2$     \\ 
\raisebox{0pt}[10pt][7pt]{$\mathrm{M}_{\mathrm{W}}$}      & $1.34^{+0.50}_{-1.34}$ &
$[0.67]^2$ & = & $[0.47]^2$ +  $[0.47]^2$ +  $[0.02]^2$ +  $[0.00]^2$     \\ 
\hline
\end{tabular}
\end{center}
\end{table}

\begin{figure}
\centering
\mbox{%
\epsfig{file=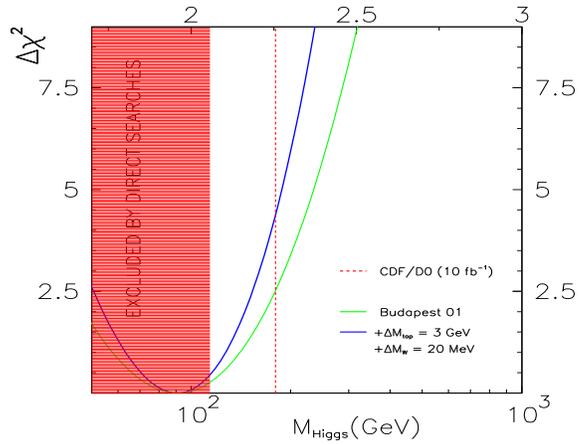 % this is the name of your figure file
        ,height=6.5cm  % this is the height of the figure (optional)
        ,width=8cm   % this is the width of the figure (optional)
       }%
}
\caption{ $\Delta\chi^2 = \chi^2 - \chi^2_{\mathrm{min}}$ vs. \MH~curve. Different 
cases are considered: the present situation and the future situation when 
$\Delta$\MW~is measured with a precision of 20~MeV  
and $\Delta$\mt=3~GeV.
The band shows the limit from direct searches, and the discontinous line the 
expected limit from TEVATRON with $10~\mathrm{fb}^{-1}$.} % the figure caption
\label{fig:chi2mh}      % the (optional) label to refer to in the text
\end{figure}

\subsection{\bf What's next?}\label{subsec:future}

LHC and its multipurpose detectors (ATLAS/CMS) are the ideal laboratory 
to disentangle the mystery of mass generation in the MSM. It's therefore 
interesting to evaluate what could be the situation of the indirect 
determination of the Higgs mass when LHC starts sometime in 2006.

TEVATRON has started its RUN II program and CDF/D0 expect to collect 
a significant amount of luminosity during the coming years. In 
figure~\ref{fig:chi2mh} it is shown the expected direct limit on \MH~with 
$10~\mathrm{fb}^{-1}$. It's also shown what would be the improvement 
in the indirect determination when CDF/D0 are able to measure the 
top mass with an uncertainty of 3~GeV and improve the world average of 
the W mass to give an uncertainty of 20~MeV.
 
Either the Higgs particle is found relatively soon (it may be that LEP has 
already seen the first hints~\cite{LEPHIGGS} of it), or the MSM will be in real trouble 
to describe the precision measurements.

\section{Conclusions and outlook}

The measurements performed at LEP and SLC have substantially improved 
the precision of the tests of the MSM, at the level of \cal{O}(0.1\%).
The effects of pure weak corrections are visible with a significance 
of about five standard deviations from \Z0~observables and about 
twelve standard deviations from the W-boson mass.

The top mass predicted by the MSM fits, (\mt=$180^{+11}_{-9}$~GeV) is
in very good agreement with the direct measurement 
(\mt=$174.3 \pm 5.1$~GeV) and of similar precision.

The W-boson mass predicted by the MSM fits, ($\mathrm{M}_{\mathrm{W}}=80.375\pm0.022~\mathrm{GeV}$) 
is compatible (-1.9$\sigma$) with the direct measurement 
($\mathrm{M}_{\mathrm{W}}=80.451\pm0.033~\mathrm{GeV}$).

The mass of the Higgs boson is predicted to be low,

\noindent
\bea
\log(\mathrm{M}_{\mathrm{H}}/\mathrm{GeV}) = & 1.96 \pm 0.19       & 
                                   (\mathrm{M}_{\mathrm{H}} = 91^{+49}_{-33}~\mathrm{GeV}) \nonumber \\ 
\mathrm{M}_{\mathrm{H}} &  < & 196~\mathrm{GeV}~@ 95 \%~\mathrm{C.L.}  \nonumber 
\eea

All measurements are internally consistent with the predictions of the MSM and with 
a very low value of the Higgs mass (lower than the limit from direct searches), with the exception 
of the hadronic \Z0 asymmetries that prefers a somehow larger value of the Higgs mass.

%\noindent This uncertainty is reduced to $\Delta(\log(\mathrm{M}_{\mathrm{H}}/\mathrm{GeV})) \sim 0.23$ when
%the uncertainty from 
%$\Delta\alpha$ is negligible, and will be further reduced to  
%$\Delta(\log(\mathrm{M}_{\mathrm{H}}/\mathrm{GeV})) \sim 0.15$ when \mt~is known with a 2~GeV precision 
%and \MW~is known with a 30~MeV precision.

\section*{Acknowledgments}

I would like to thank Antonio Dobado, Victoria Fonseca and all the organizing committee for the 
excellent organization of the meeting.
%I'm very grateful to Martin W. Gr$\ddot{\mathrm{u}}$newald and Gunter Quast for his help 
%in the preparation of the numbers and plots shown in this paper. I also thank
%Guenther Dissertori for reading the paper and giving constructive criticisms.

%This is where one places acknowledgments for funding
%bodies etc.  Note that there are no section numbers for
%the Acknowledgments, Appendix or References.

%\section*{Appendix}


\section*{References}
\begin{thebibliography}{99}

\bibitem{PDG}The Particle Data Group, D. E.~Groom {\it et al}, \Journal{\EPJ}{15}{1}{2000}.

\bibitem{STEINHAUSSER} M. Steinhauser, \Journal{\PLB}{429}{158}{1998}.

\bibitem{PIETRZYK}H. Burkhardt and B. Pietrzyk, LAPP-EXP 2001-03 accepted by Physics Letters B.

\bibitem{DAVIER}M. Davier and A. H$\ddot{\mathrm{o}}$cker, \Journal{\PLB}{419}{419}{1998}.

\bibitem{NEWALPHA}J.H. K$\ddot{\mathrm{u}}$hn and M. Steinhauser, \Journal{\PLB}{437}{425}{1998}. \\
                  F. Jegerlehner, Proceedings of Fourth International Symposium on Radiative Corrections,
                  Barcelona, September 98, pag. 75. \\
                  J. Erler, \Journal{\PRD}{59}{054008}{1999}.\\
                  A. D. Martin, J. Outhwaite and M. G. Ryskin, \Journal{\PLB}{492}{69}{2000}. 

\bibitem{BESS} The BESS Collaboration, J. Z. Bai, hep-exp/0102003.

%\bibitem{WARD}W. Plazcek, {\it these proceedings}.

\bibitem{TOPAZ0}G. Passarino {\it et al}, hep-ph/9804211.

\bibitem{ZFITTER}D. Bardin {\it et al}, hep-ph/9412201.

%\bibitem{ALPHA3}S. Jadach {\it et al}, \Journal{\PLB}{257}{173}{1991}. \\
%                M. Skrzypek {\it et al}, \Journal{\APP}{23}{135}{1992}. \\
%                G. Montagna {\it et al}, \Journal{\PLB}{406}{243}{1997}. 

%\bibitem{QCD4LOOP}S. Larin {\it et al}, \Journal{\PLB}{400}{379}{1997}. \\
%                  K.G. Chetyrkin, \Journal{\PLB}{404}{161}{1997}. \\
%                  S. Larin {\it et al}, \Journal{\PLB}{405}{327}{1997}. \\
%                  K.G. Chetyrkin {\it et al}, \Journal{\PRL}{79}{2184}{1997}. \\
%                  K.G. Chetyrkin {\it et al}, \Journal{\NPB}{510}{61}{1998}. 

%\bibitem{QCDEW}A. Czarnecki {\it et al}, \Journal{\PRL}{77}{3955}{1996}. \\
%               R. Harlander {\it et al}, \Journal{\PLB}{426}{125}{1998}.

%\bibitem{DEGRASSI}G. Degrassi {\it et al}, \Journal{\PLB}{383}{219}{1996}. \\
%                  G. Degrassi {\it et al}, \Journal{\PLB}{394}{188}{1997}. \\
%                  G. Degrassi {\it et al}, \Journal{\PLB}{418}{209}{1997}. 

\bibitem{YR95}CERN Yellow Report 95-03, Geneva, 31 March 1995, eds. D.~Bardin, W.~Hollik and 
              G.~Passarino.

\bibitem{LEPHIGGS} F. Gianotti, ``Searches for new particles at colliders'', International Europhysics Conference
                                on High Energy Physics, Budapest, July 12-18, 2001.

%\bibitem{LANCON}E. Lan\c{c}on, {\it these proceedings}.

%\bibitem{SIRLIN}P. Gambino and A. Sirlin, \Journal{\PRL}{73}{621}{1994}.

%\bibitem{NUTEV}K. MacFarland, talk presented at the XXXIIIth Rencontres de Moriond, Les Arcs,
%               France, 15-21 March, 1998.

\end{thebibliography}
\end{document}